\input psfig
\documentstyle[11pt]{article}
\def\Journal#1#2#3#4{{#1} {\bf #2}, #3 (#4)}
\def\NCA{{\em Nuovo Cimento} A}

\def\NPB{{\em Nucl. Phys.} B}
\def\PLA{{\em Phys. Lett.}  A}
\def\PLB{{\em Phys. Lett.}  B}
\def\PRL{\em Phys. Rev. Lett.}
\def\PR{\em Phys. Rev.}
\def\PRA{{\em Phys. Rev.} A}
\def\PRB{{\em Phys. Rev.} B}
\def\PRD{{\em Phys. Rev.} D}

\begin{document}

\title{On the Divergence of  Perturbation Theory.  \\
         Steps Towards a Convergent Series.}
\author{ Sergio A. Pernice, \\
Departmant of Physics and Astronomy, \\
University of Rochester,\\
Rochester, New York, 14627.  \\
\\
Gerardo Oleaga, \\
Departamento de Matematica Aplicada,\\
Facultad de Matematicas,\\
Universidad Complutense,\\
28040 Madrid, Spain.}
\maketitle

e-mail: sergio@charm.pas.rochester.edu

\begin{abstract}
The mechanism underlying the divergence of perturbation theory is exposed.
This is done through a detailed study of  the violation of the hypothesis of the
Dominated Convergence Theorem of Lebesgue using familiar techniques of
Quantum Field Theory. That theorem governs the validity
(or lack of it) of the formal manipulations done to generate the perturbative
series in the functional integral formalism. The aspects of the perturbative series
that need to be modified to obtain a convergent series are presented. Useful
tools for a practical implementation of these modifications are developed.
Some resummation methods are analyzed in the light of the above mentioned
mechanism.
\end{abstract}

\newpage

\section{Introduction}\label{sec:intr}

A typical quantity to analyze the nature of the
perturbative expansion in Quantum Field Theory is the partition function
\begin{equation} 
Z ( \lambda ) = {1 \over  Z_0} \int  \left[ {\rm d} \phi  \right] 
e^{- S  \left[ \phi \right]}
\label{eq:partfunc} 
\end{equation} 
with
 \begin{equation} 
S  \left[ \phi \right] = \int   {\rm d^d} x  \left[ {1 \over 2}  \left( \partial_{\mu} \phi \right)^2
+ {1 \over 2}  m^2 \phi^2 + {\lambda \over 4} \phi^4 \right].
\label{eq:action} 
\end{equation} 
The normalization  factor  $1/Z_0$ is the partition function of the free field
($Z \rightarrow 1 $ when $ \lambda \rightarrow 0  $). The analysis of the
perturbative expansion of any Green's function goes along similar lines as 
in the case of $Z$. In the example above we consider a scalar field 
theory for simplicity.

The traditional argument for understanding the divergent nature of the
perturbative expansion can be traced back to Dyson~\cite{dy}. Although the
form was different,  the content of his argument is captured by the 
following statement:

``If the perturbative series were to converge to the exact result, the function being 
expanded would be analytic in  $\lambda$ at $\lambda = 0$.
But the function (Z for example) is not analytic in $\lambda$ at that
value. Therefore, as a function of $\lambda$, the perturbative series
is either divergent or converges to the wrong answer."

Estimations of the large order behavior of the coefficients
of the perturbative series showed that the first possibility is the one
actually realized \cite{lo,bw}. 
That Z, as a function of $\lambda$, is not analytic at $\lambda = 0$,
can be guessed by simply noting that if in its functional 
integral representation (Eq.~(\ref{eq:partfunc})) we make the real part of
 $\lambda$ negative, the integral diverges.  
In fact, there is a branch cut in the first Riemann sheet that can be 
chosen to lie along the negative real axis, extending from 
$\lambda = - \infty$ to $ \lambda  = 0$ \cite{bw0,bs}.

The above argument is very powerful and extends to the perturbative
series of almost all other nontrivial field theories. It
has also motivated a series of very important calculations of the 
large order behavior of the perturbative coefficients \cite{lo}, general analysis
on the structure of field theories \cite{tHooft}, as well as improvements over
perturbative computations of different physical quantities \cite{lo}.

For all its power, it is fair to say that the argument,
as almost any other {\it reductio ad absurdum} type of argument,
fails to point towards a solution of the problem of divergence. 
It is only through the indirect formalism of Borel transforms that
questions of recovery of the full theory from its perturbative
series can be discussed \cite{bs,gh,zinn}.

In this paper an alternative way of understanding the divergent
nature of the perturbative series is presented. This way of 
understanding the problem complements the traditional argument 
briefly described above, hopefully illuminating aspects that
the traditional approach leaves obscure. In particular,
as we will see, the arguments in this paper point directly
towards the aspects of the perturbative series that need to
be modified to achieve a convergent series. It is hoped that the 
way of understanding the problem presented here will help to provide
new insights into the urgent problem  of extracting non-perturbative
information out of Quantum Field Theories.

In section~\ref{sec:leb} we develop our analysis of the divergence of
perturbation theory. In  Sec.~\ref{sec:converg} we point out the ingredients
that, according to the analysis of  Sec.~\ref{sec:leb}, a modification
of perturbation theory would need to achieve convergence. We also present
a remarkable formula~(\ref{win}) that allows us to implement such modifications
in terms of Gaussian integrals, paving the way to the application of this
convergent modified perturbative series to Quantum Field Theories. The proof
of the properties of the function~(\ref{win}) is done in Appendix 1.
In section~\ref{Improvement} we analyze recent work on the convergence of
various optimized expansions~[12-19]  in terms of the ideas
presented here.
In  Sec.~\ref{sec:concl} we summarize our results and mention directions
of the work currently in preparation. Finally, in Appendix 2, we apply
the ideas of this paper in a simple but illuminating example for which we
actually develop a convergent series by modifying the aspects of the
perturbative series pointed out by our analysis as the source of divergence.

\section{Lebesgue's Dominated Convergence Theorem and Perturbation
Theory}\label{sec:leb}

\subsection{The wrong step in perturbation theory}\label{sec:wrong}

Although the notation will not always be explicit, we work in an Euclidean 
space of dimension smaller than 4 and in a finite volume.

Let's remember how the perturbative series is generated in the 
functional integral formalism for a quantity like $Z$: 
\begin{eqnarray}\label{eq:pertZ}
Z ( \lambda ) & = & \int  \left[ {\rm d} \phi  \right] e^{-
\int   {\rm d^d} x  \left[ {1 \over 2}  \left( \partial_{\mu} \phi \right)^2
+ {1 \over 2}  m^2 \phi^2  \right]  -  {\lambda \over 4}  \int   {\rm d^d} x \phi^4 }  \\
\label{eq:first}
& = &  \int  \left[ {\rm d} \phi  \right] \sum_{n=0}^{\infty} {\left( -1 \right)^n \over n!}  
\left( {\lambda \over 4}   \int   {\rm d^d} x \phi^4   \right)^n  e^{- \int   {\rm d^d} x 
 \left[ {1 \over 2}  \left( \partial_{\mu}  \phi \right)^2 + {1 \over 2}  m^2 \phi^2  \right]   }  \\
\label{eq:second}
& = &  \sum_{n=0}^{\infty} \int  \left[ {\rm d} \phi  \right]  {\left( -1 \right)^n \over n!}  
\left( {\lambda \over 4}   \int   {\rm d^d} x \phi^4   \right)^n  e^{- \int   {\rm d^d} x 
 \left[ {1 \over 2}  \left( \partial_{\mu}  \phi \right)^2 + {1 \over 2}  m^2 \phi^2  \right]   } 
\end{eqnarray}

The final sum is in practice truncated at some finite order $N$.
The functional integrals that give the contribution of every
order $n$ are calculated using Wick's theorem and Feynman's 
diagram techniques with the corresponding renormalization.

We see then that the generation of the perturbative series in the
functional integral formalism is a two step process. First  
(\ref{eq:first}) the integrand is expanded
in powers of the coupling constant, and then  (\ref{eq:second}) the sum is 
interchanged with the integral\footnote{ In this paper  we will often use the
familiar word ``integrand" to refer to  $e^{-S}$ or any functional inside the
functional integration symbol. It would be more precise to preserve this word
for $e^{-S_{\rm Int}}$ in the measure defined by the free field. The terminology
used here is, however, common practice in the Quantum Field Theory
literature and also helps to emphasize the similarities with the intuitive finite
dimensional case presented below. }

It will be convenient to have a simpler example in which the arguments
of this paper become very transparent. Consider the simple integral 
\begin{eqnarray}\label{eq:pertsimple}
z ( \lambda ) & = & { 1 \over \sqrt{\pi} }  \int_{- \infty}^{\infty} {\rm d} x
e^{- \left( x^2 + {\lambda \over 4} x^4 \right)}
\end{eqnarray}
and its corresponding  perturbative expansion
\begin{eqnarray}\label{eq:pertsimple1}
z ( \lambda ) & = & { 1 \over \sqrt{\pi} }   \int_{- \infty}^{\infty} {\rm d} x  \sum_{n=0}^{\infty} 
  {\left( -1 \right)^n \over n! } \left( {\lambda \over 4} x^4 \right)^n    e^{- x^2   } \\
\label{eq:pertsimple2}
& = & { 1 \over \sqrt{\pi} }   \sum_{n=0}^{\infty}   \int_{- \infty}^{\infty} {\rm d} x 
{\left( -1 \right)^n \over n! } \left( {\lambda \over 4} x^4 \right)^n    e^{- x^2   } \\
\label{eq:simplecoef}
& \equiv &  \sum_{n=0}^{\infty}  \left( -1 \right)^n c_n \  \lambda^n.
\end{eqnarray}
This simple integral has been used many times in the past  as a paradigmatic 
example of the divergence of perturbation theory \cite{zinn}. It  is then specially
suited for a comparison between the traditional arguments and the ones
presented  in this paper.

Again we see the two step process to generate the perturbative
series. First the integrand is expanded in powers of  $\lambda$~(\ref{eq:pertsimple1})
and then the sum is interchanged with the integral~(\ref{eq:pertsimple2}) .
In this simple example the perturbative coefficients can be 
calculated exactly for arbitrary $n$. In the large $n$ limit they become:
\begin{equation}\label{eq:simplecoeflargen}
c_n \sim  {\sqrt{2} \over 2 \pi }  \left( n -1 \right)!  \quad  {\rm when} 
\quad  n \rightarrow \infty.
\end{equation}

With such factorial behavior, the series diverges for all $ \lambda$ 
different from zero as is well known. On the other hand the function $z (
\lambda )$, as defined in  Eq.~(\ref{eq:pertsimple}),  gives a well defined
positive real number for every positive  real $\lambda$. Therefore one or 
both of the two steps done to generate the perturbative series must be wrong.

Similarly, in the functional integral case normalized with respect to the free
field (\ref{eq:partfunc}), $Z$ is a well defined number while its
perturbative series diverges. One or both of the two steps must be wrong.

The first step, the expansion of the integrand in powers of $ \lambda$,
is clearly correct. As the integrand (not the integral!) is analytic 
in $ \lambda$ for
every finite $ \lambda$, the expansion  merely corresponds to a Taylor
series. The second step, the interchange of sum and integral,  must therefore
be the wrong one.

The next obvious step is then to recall the theorems that govern
the interchange between sums and integrals, to understand in detail
why this is wrong in our case. The most powerful theorem in this
respect is the well known theorem of Dominated Convergence of
Lebesgue. In a simplified version, enough for our purposes, it
says the following:

\begin{quote}
Let $f_N$ be a sequence of integrable functions that converge
pointwisely to a function $f$
\begin{equation}\label{eq:pointconv}
f_N \longrightarrow f   \quad  {\rm as} \quad  N  \rightarrow \infty
\end{equation}
and bounded in absolute value  by a positive integrable function $h$
(dominated)
\begin{equation}\label{eq:bound}
 |f_N|  \leq h \ ,\quad \forall N.
\end{equation}
Then, it is true that
\begin{equation}\label{eq:thesis}
 \lim_{N \rightarrow \infty} \int  f_N  =  \int  \lim_{N \rightarrow \infty}   f_N 
= \int f .
\end{equation}
\end{quote}
As a special case, if the convergence~(\ref{eq:pointconv}) is uniform and
the measure of integration is finite,  then the interchange is also valid. It
should be emphasized that Lebesgue's theorem follows from the axioms of
abstract measure theory. Therefore if the problem under consideration
involves a well defined measure, as is the case for the Quantum Field
Theories considered here~\cite{glja}, the theorem holds.

In our case we can write formally\footnote{See previous footnote.},
\begin{equation}\label{eq:fNfunctint}
f_N \left[ \phi (x) \right]=  {1 \over Z_0} \sum_{n=0}^{N}  
{\left( -1 \right)^n \over n!}  
\left( {\lambda \over 4}   \int   {\rm d^d} x \phi^4   \right)^n  e^{- \int   {\rm d^d} x 
 \left[ {1 \over 2}  \left( \partial_{\mu}  \phi \right)^2 + {1 \over 2}  m^2 \phi^2  \right]   }
\end{equation}
for the functional integral case, and
\begin{equation}\label{eq:fNsimpltint}
f_N (x)  = { 1 \over \sqrt{\pi} } \sum_{n=0}^{N}  {\left( -1 \right)^n \over n!}  
\left( {\lambda \over 4}   x^4   \right)^n  e^{- x^2 }
\end{equation}
for the simple integral example.

One important aspect of the dominated convergence theorem approach
to analyze the divergence of perturbation theory is that it
focuses on the integrands, objects relatively simple to analyze.
On the contrary, the analyticity  approach briefly described in the
introduction focuses on the integrals, that are much more difficult
to analyze. So, before we try to understand what aspects of the
dominated convergence theorem fail in our case, let's see the
``phenomena" (the integrand) for the intuitive simple example.
\begin{figure}[htp]
\hbox to \hsize{\hss\psfig{figure=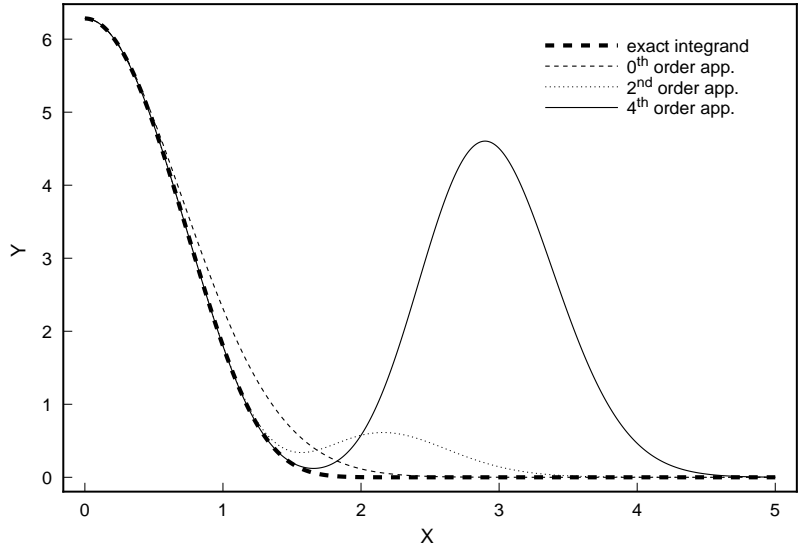,width=0.9\hsize}\hss}
\caption{Exact integrand,  ${\rm zero}^{{\rm th}}$, second and fourth perturbative
approximations. $\lambda = 1$.}   
\label{fig1}
\end{figure}
In figure ~\ref{fig1}, the exact integrand, together with some  perturbative
approximations,
are displayed. We can appreciate the way in which the successive
approximations behave. For small $x$, and up to some
critical value that we call $x_{c,N}$, where the subindex $c$ stands
for {\it critical} while the subindex $N$ indicates that this value changes with the 
order, the perturbative integrands approximate very well the exact integrand.
Even more, $x_{c,N}$ grows with $N$. But for $x$ bigger that  $x_{c,N}$ a  ``bump''
begins to emerge. The height of these bumps, as we will see in detail shortly,
grows factorially with the order, while the width remains approximately constant.
So, the larger the order in perturbation theory, the larger the region in which
the perturbative integrands approximate very well the exact integrand, but
the stronger the upcoming deviation. As we will see shortly, it is precisely
this deviation that is responsible for the divergence of the perturbative series
and the famous factorial growth. We will also see that  an exactly  analogous
phenomena  happens in the functional integral case and is again the responsible
for the divergence of the perturbative series.

Returning to the problem of  understanding the aspect of the dominated convergence
theorem that fail in the perturbative series we will now show that the sequence
of integrands  of Eq.~(\ref{eq:fNfunctint})  and  Eq.~(\ref{eq:fNsimpltint})
converge respectively to the exact integrands
\begin{equation}\label{eq:exIntd}
F = {1 \over Z_0} \ e^{- \int   {\rm d^d} x  \left[ {1 \over 2}  
\left( \partial_{\mu} \phi \right)^2
+ {1 \over 2}  m^2 \phi^2  \right]  -  {\lambda \over 4}  \int   {\rm d^d} x \phi^4 } 
\end{equation}
and
\begin{equation}\label{eq:exintd}
f = {1 \over \sqrt{\pi}}  e^{- \left( x^2 + {\lambda \over 4} x^4 \right)}
\end{equation}
but {\it not} in a {\it dominated} way. This is, there is no positive integrable
function $h$ satisfying the property of  Eq.~(\ref{eq:bound}).

\subsection{Failure of domination in the simple example}\label{sec:simpleint}

\begin{figure}[htp]
\hbox to \hsize{\hss\psfig{figure=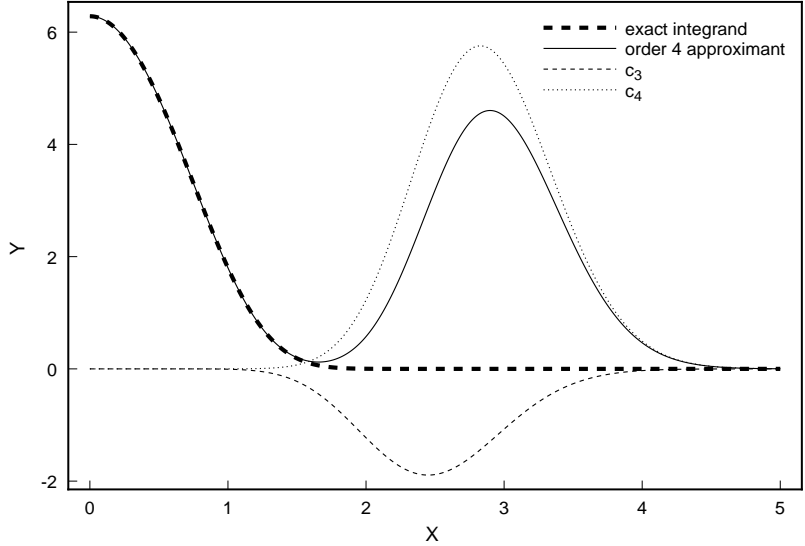,width=0.9\hsize}\hss}
\caption{Exact integrand, and fourth perturbative approximation together with
the third, and fourth terms. $\lambda = 1$.}   
\label{fig2}
\end{figure}

That  the sequence of integrands of Eq.~(\ref{eq:fNfunctint})  and
 Eq.~(\ref{eq:fNsimpltint}) converge respectively to the exact integrands
 (\ref{eq:exIntd})    and   (\ref{eq:exintd})
is obvious, since, as mentioned before, for finite $\lambda$ they are
analytic functions of $\lambda$  and so their Taylor
expansion converge (at least for finite field strength). To see the failure of  the domination hypothesis it is convenient to analyze
the ``shape" of every term of $f_N$. Namely, for the field theory case, 
\begin{equation}\label{eq:cnFT}
c_n   \left[ \phi (x) \right] \equiv  {1 \over Z_0} {\left( -1 \right)^n \over n!} 
 {1  \over 4^n}
\left(  \lambda \int   {\rm d^d} x \phi^4   \right)^n  e^{- \int   {\rm d^d} x 
 \left[ {1 \over 2}  \left( \partial_{\mu}  \phi \right)^2 + {1 \over 2}  m^2 \phi^2  \right]   }
\end{equation}
while for the simple integrand
\begin{equation}\label{eq:cnSI}
c_n  (x)  =  {1 \over \sqrt{\pi}} {\left( -1 \right)^n \over n!}   \left( {\lambda \over 4} \right)^n   x^{4n} 
 e^{- x^2 }.
\end{equation}
In this section we analyze the failure of the domination hypothesis for the
simple example~(\ref{eq:pertsimple}) because, as it turns out, it is
remarkably similar to the Quantum Field Theory example analyzed in the
next section.
In Fig.~\ref{fig2} we can inspect the functions $c_3 (x)$ and $c_4(x)$ for
$\lambda = 1$ corresponding to the
simple integrand case that we analyze first. The maximum of $c_n (x)$ is reached
at 
\begin{equation}\label{eq:xmax}
x_{{\rm max}} = \pm (2 n)^{1/2}.
\end{equation}
There, for large $n$, the function takes the value 
\begin{equation}\label{eq:cnmax}
c_n  (x_{{\rm max}}) =  {1 \over 2 \pi^{3/2}}  (-1)^n (n-1)! \ \lambda^n .
\end{equation}
 On the other hand, the width remains constant as n increases
as can be seen by a Gaussian approximation around the maximum 
$x_{{\rm max}} = (2 n)^{1/2}$:
\begin{equation}\label{gaussapp}
c_n (x) \approx  {1 \over 2 \pi^{3/2}}  (-1)^n (n-1)! \  {\rm exp}  
\left[ -2 (x - (2n)^{1/2})^2  \right] \ \lambda^n .
\end{equation}
The integration of this Gaussian approximation gives, for large $n$
\begin{equation}\label{gaussapp2}
\int {\rm d} x \ c_n (x)  \approx  {1 \over 2} { \sqrt{2} \over 2 \pi}  (-1)^n (n-1)!  \ 
\lambda^n 
\end{equation}
in accordance with Eq.~(\ref{eq:simplecoeflargen}) if we take into account
the factor of  2 coming from the two maxima $\pm  (2 n)^{1/2}$.

The mechanism of convergence of the $f_N $'s to $f$  becomes clear now.
The $f_N $'s are made out of  a pure Gaussian (the ``free" term) plus
``bumps" (the perturbative corrections) oscillating in sign (see
Fig.~\ref{fig2}). The maxima of these bumps grows factorially with the order
while their width remain approximately constant (more specifically, the 
Gaussian approximation around the maxima (Eq.~(\ref{gaussapp})), that becomes
exact
when the order goes to infinity, has a variance independent of the order). For
fixed $N$, and for $x$ smaller than  a certain value, the bumps delicately
``almost" cancel each other, leaving only a small remnant that modifies the 
free integrand into the interacting one. However, for $x$ larger than that
value, the  last bump begins to emerge and, being the last, does not have the
next one to get cancelled (in the $N \rightarrow \infty$ limit, there is no 
last bump and the convergence is achieved for every x). Consequently, after 
a certain value $x_{c,N}$, the function $f_N$ deviates strongly from $f$ and
is governed by the uncanceled $N^{{\rm th}}$ bump only, of height  
proportional to $(N-1)!$ and finite variance. This is so because, since 
the height of the bump grows factorially with the order, for $N$ large enough
the last bump is far greater than all the previous ones, so it remains almost
completely uncanceled. Since the variance of the bumps is independent
of the order, this  means that for every finite order, there is a region of
{\it finite measure} in which the perturbative integrand is of the order
of the height of the last bump. In figure ~\ref{fig2} we can 
see how the function $c_4 (x)$ is left almost completely uncanceled by 
$c_3 (x)$ and dominates the deviation of $f_4$ from $f$.

That $x_{c,N}$ (the value of $|x|$ up to which the perturbative integrand very
accurately approximate the exact one) grows with $N$, going to infinity when
$N \rightarrow \infty$, is a simple consequence of  Taylor's theorem applied
to the analytic function $e^{- \lambda x^4 /4}$.

The above analysis makes clear the failure of domination of the sequence
of  Eq.~(\ref{eq:fNsimpltint})  towards $f$ (Eq.~(\ref{eq:exintd})). Indeed, any positive
 function $h(x)$ with the property
\begin{equation}\label{eq:domSI}
|f_N (x)|  \leq h (x)\ ,\quad \forall N
\end{equation}
fails to be integrable, since it has to ``cover" the bump, whose area grows factorially
with $N$. So, although the sequence of  $f_N (x)$'s converges to $f(x)$, the
convergence is not dominated as we wanted to show.

 Eq.~(\ref{gaussapp2}), together with the above comments, indicate
that the same reason for which the sequence of  integrands (\ref{eq:cnSI}) 
fails to be dominated, is the one that produces the factorial growth in
the perturbative series.

In the Field Theory case, although we can not rely on figures 
like~\ref{fig1} and \ref{fig2} to guide our intuition, as we will show now,
the analogy with the simple integral example is so close that the interpretation
is equally transparent.

\subsection{Failure of domination in Quantum Field Theory}\label{sec:QM}

For  Quantum Field Theory, as for the simple example analyzed above,
it is convenient to consider every term $c_n [ \phi (x) ]$ 
(Eq.~\ref{eq:cnFT}) of the perturbative approximation  $f_N$ 
(Eq.~\ref{eq:fNfunctint}) of the exact  integrand (Eq.~\ref{eq:exIntd}),
\begin{equation}\label{eq:cnFT2}
c_n   \left[ \phi (x) \right]  =  {1 \over Z_0}{\left( -1 \right)^n \over n!}  
 e^{- \int   {\rm d^d} x   \left[ {1 \over 2}   \left(  \partial_{\mu} \phi \right)^2 +
 {1 \over 2}  m^2 \phi^2  \right]   +  n \ln{\left[  \left(  \lambda/4 \right)  
\int   {\rm d^d} x \ \phi^4   \right] } }
\end{equation}
where we have written the $n^{{\rm th}}$ power of the interaction in exponential
form.  The mathematical analysis below follows closely the discussions in
chapter 38 Ref.~\cite{zinn}. Although the problem treated there is 
different from the one treated here, many techniques used in~\cite{zinn}
can be directly borrowed here.

For $n$ large enough, the analysis of its ``shape" reduces to the familiar
procedure of finding its maxima, as in the case of the simple integrand. The
equation determining the maxima of  $c_n [ \phi (x) ]$ is the equation that  minimizes the exponent, and can be thought of as the equation of motion of the
effective action
\begin{equation}\label{eq. effactQM}
S \left[  \phi \right] = \int   {\rm d^d} x   \left[ {1 \over 2}  \left(  \partial_{\mu} 
\phi \right)^2 + {1 \over 2}  m^2 \phi^2  \right]   -  n \ln{ \left[  
{\lambda \over 4}  \int   {\rm d^d} x  \ \phi^4   \right] },
\end{equation}
which is
\begin{equation}\label{eq:eqQM}
- \nabla^2 \phi + m^2 \phi - { 4 n \over  \int   {\rm d^d} x \phi^4 }  \phi^3 = 0.
\end{equation}
Making the change of variables
\begin{equation}\label{eq:chvarQM1}
\phi ( x ) = m  \left({ \int   {\rm d^d} x \phi^4 \over 4n} \right)^{1/2} \varphi ( m x )  =
 m^{ {\rm d}/2 -1} \left(  {4n \over     \int   {\rm d^d} u \varphi^4 (u) }
\right)^{1/2} \varphi ( m x ), 
\end{equation}
$ \varphi $ satisfies the equation
\begin{equation}\label{eq:numeqQM}
 -  \nabla^2 \varphi (u) + \varphi (u) -  \varphi^3 (u) = 0 \quad ,
\qquad u \equiv m x.
\end{equation}
This equation corresponds to the instanton equation of the negative mass
$\lambda \phi^4$ theory. The analysis of their  solutions can be found in many
places. We are interested in the solutions with minimal, finite action. For
these solutions, in the infinite volume limit, scaling arguments provide very
interesting information. We mentioned at the beginning of 
section~\ref{sec:wrong} that we work in a finite volume. However, if the
volume is large enough, the following infinite volume arguments remain
valid up to errors that go to zero exponentially fast when the volume
goes to infinity.

Since the solution $\phi_{{\rm max}} (x)$ (the subindex ``max" indicates
that, in functional space, $c_n   \left[ \phi (x) \right]$ reaches its maximum
at $\phi_{{\rm max}} (x)$,  this should not be confused with the fact that  the
the action (\ref{eq. effactQM}) reaches its {\it minimum}  there) is a minimum
of the action (\ref{eq. effactQM}), given an arbitrary constant $\alpha$,  
$S \left[  \alpha \phi_{{\rm max}} (x) \right]$ should have a minimum at $\alpha  = 1$ \cite{zinn,derr}.
This implies the equation
\begin{equation}\label{Salphphi1}
\int   {\rm d^d} x   \left(  \partial_{\mu}   \phi_{{\rm max}} \right)^2 + m^2  \int   {\rm d^d} x    \phi_{{\rm max}}^2  
- 4n = 0. 
\end{equation}
Similarly, $S \left[    \phi_{{\rm max}} (\alpha x) \right]$ should also have a minima 
at  $\alpha = 1$, implying,
\begin{equation}\label{Salphphi2}
{ \left( 2 - d \right) \over d}    \int   {\rm d^d} x   \left(  \partial_{\mu}   \phi_{{\rm max}} \right)^2 -
  m^2  \int   {\rm d^d} x    \phi_{{\rm max}}^2  +  2  n = 0. 
\end{equation}
Solving the system of equations  (\ref{Salphphi1}) and (\ref{Salphphi2}) we obtain,
\begin{eqnarray}\label{solusysteq1}
\int   {\rm d^d} x   \left(  \partial_{\mu}   \phi_{{\rm max}} \right)^2 &=& n \ d  \\  \label{solusysteq2}
m^2  \int   {\rm d^d} x    \phi_{{\rm max}}^2 &=& n  \left( 4 - d \right).
\end{eqnarray}
From which we conclude in particular
\begin{equation}\label{eq. effactQMFree}
 \int   {\rm d^d} x   \left[ {1 \over 2}  \left(  \partial_{\mu} 
 \phi_{{\rm max}} \right)^2 + {1 \over 2}  m^2  \phi_{{\rm max}}^2  \right]  = 2n,
\end{equation}
independent of the dimension. The relations (\ref{solusysteq1}) and (\ref{solusysteq2})
can be explicitly checked in the case $d=1$ (Quantum Mechanics), in which the
solutions to eq.(\ref{eq:eqQM}) are known analytically. They are
\begin{equation}\label{d=1sol}
\phi_{{\rm max}}^{{\rm d} =1} (t) =  \left( {3n \over 2 m} \right)^{1/2} {1 \over 
\cosh{ \left[ m \left( t - t_0 \right) \right]} }
\end{equation}
giving
\begin{eqnarray}\label{solusysteq1D1}
\int   {\rm d} t   \left(    \dot{ \phi}_{{\rm max}}^{{\rm d} =1} \right)^2 &=& n   \\  \label{solusysteq2D1}
m^2  \int   {\rm d} t    \left( \phi_{{\rm max}}^{{\rm d} =1} \right)^2 &=& 3 n .
\end{eqnarray}

Since $\varphi (u)$, introduced in eq.(\ref{eq:chvarQM1}) and satisfying 
eq.(\ref{eq:numeqQM}), is dimensionless (remember that $u = m x$ is also 
dimensionless), and the corresponding  $\phi_{{\rm max}} (x)$ has finite action, 
the quantity
\begin{equation}\label{Adef}
A \equiv  { 1 \over 4}  \int {\rm d^d}  u    \varphi^4  (u)
\end{equation}
is a finite, pure number greater than zero~\cite{zinn}. For the Quantum
Mechanical case 
mentioned above, $A = 4/3$. For the cases $d>1$, $A$ is not explicitly known
but, as just said, it must be a finite, positive, pure number. With the
definition~(\ref{Adef}), Eq.~(\ref{eq:chvarQM1}) becomes,
\begin{equation}\label{relphimaxvarphi}
\phi_{{\rm max}} (x) = m^{d/2 - 1}  \left( { n \over A} \right)^{1/2} \varphi  (m x)
\end{equation}
Since  $\varphi  (m x)$ satisfies the n-independent Eq.~(\ref{eq:numeqQM}), we
conclude that  the field strength of  $\phi_{{\rm max}}$ grows with the square
root of  the order $n$.

Equation (\ref{eq. effactQMFree}), together with the definition (\ref{Adef}) and
the relation  (\ref{relphimaxvarphi}), allow us to write an expression for the
action  (\ref{eq. effactQM}) at  $\phi = \phi_{{\rm max}}$,
\begin{equation}\label{Sphimax}
S  \left[ \phi_{{\rm max}} \right] = 2 n - n \ln{ \left[  {\lambda \  m^{d-4} \over A}  n^2
 \right]  }
\end{equation}
The value of  $c_n   \left[ \phi (x) \right]$ at  $\phi = \phi_{{\rm max}}$ then
becomes, for large n,
\begin{equation}\label{cnphimax}
c_n   \left[ \phi_{{\rm max}} (x) \right]  \approx  {1 \over Z_0} 
{ \left( -1 \right)^n  \over  2 \pi}
\left( n - 1 \right) !  \left( {\lambda \  m^{d-4} \over A} \right)^n.  
\end{equation}

With the change of variables
\begin{equation}\label{chvarphiq}
\phi (x) = \phi_{{\rm max}} (x)  +  m^{d/2 - 1} \phi_{\rm q} (m x),
\end{equation}
the Gaussian approximation of $c_n   \left[ \phi \right]$ around  $\phi_{{\rm max}}$ is, 
\begin{eqnarray}\label{eq:gaussapp}
c_n   \left[ \varphi (u) \right]  &  \approx     & {1 \over Z_0}
{ \left( -1 \right)^n  \over  2 \pi}
\left( n - 1 \right) !  \left( {\lambda \  m^{d-4} \over A} \right)^n \cdot  \nonumber \\ 
 \label{local}
 &  & e^{- {1\over 2}  \int  {\rm d^d} u_1  {\rm d^d} u_2  \phi_{\rm q} (u_1)  \left[ \left( - 
\nabla^2_{u_1} + 1 - 3 \varphi^2 (u_1) \right)  
\delta (u_1 - u_2)\right]   \phi_{\rm q} (u_2)  }  \cdot   \\   \label{nonlocal}
&  & e^{- {1\over 2}  \int  {\rm d^d} u_1  {\rm d^d} u_2  \phi_{\rm q} (u_1)  \left[ 
 \left( 1 / A \right) \varphi^3 (u_1)  \varphi^3 (u_2)   \right]   \phi_{\rm q} (u_2)  } 
\end{eqnarray}
where $u = mx$ and $\varphi (u)$, solution of  Eq.(\ref{eq:numeqQM}), 
is related to  $\phi_{{\rm max}}$ through Eq.(\ref{relphimaxvarphi}). This 
Gaussian approximation becomes exact in the limit  $n \rightarrow \infty$. 

The second derivative operator, that we call $D$, is then,
\begin{equation}\label{D}
D  = D_{{\rm local}}  +  D_{{\rm non-local}}
\end{equation}
with
\begin{equation}\label{Dlocal}
D_{{\rm local}} =  - \nabla^2 + 1 - 3 \varphi^2  
\end{equation}
and
\begin{equation}\label{Dnonlocal}
D_{{\rm non-local}} =  { 1 \over A}  |v> <v| , \quad   {\rm with}\quad
  < u | v > = \varphi^3 (u)
\end{equation}
and $A$ given in Eq.(\ref{Adef}).

The operator $D_{{\rm local}}$ is well known (see for example~\cite{zinn}). 
It has $d$ eigenvectors 
$| 0_{\mu} >$ with zero eigenvalues given by
\begin{equation}\label{zeroeig}
< u | 0_{\mu} > = {\partial \over \partial u^{\mu}}{ \varphi (u) }.
\end{equation}
These vectors are also zero-eigenvectors of $D$, as can be seen by noting
that  $|v>$ is orthogonal to them,
\begin{equation}\label{orthv0}
< v |  0_{\mu} >  = 0.
\end{equation}
They reflect the translation invariance of the action  (\ref{eq. effactQM}).

$D_{{\rm local}}$ is also known to have one and only one negative eigenvector.
The proof of this fact  given in Appendix  38 of reference~\cite{zinn},  that uses
Sobolev inequalities, can be repeated line by line to prove that, on the
contrary, $D$ is a positive semi-definite operator,   
\begin{equation}\label{Dpossemidef}
D \ge 0
\end{equation}
in the operator sense.

Projecting out the d-dimensional eigenspace of eigenvalue zero, the resulting
operator, that we call $D'$, is positive definite.
\begin{equation}\label{Dposdef}
D' =  D'_{{\rm local}}  + D_{{\rm non-local}} > 0
\end{equation}
This equation explicitly states that the projection over the strictly positive 
eigenvectors modifies only $D_{{\rm local}}$. The non-local part, as we saw,
is a projector orthogonal to the zero modes and is therefore not modified under 
that operation.

Equations  (\ref{Dpossemidef})  and   (\ref{Dposdef}) suggest that the operator
$D$, with the corresponding  renormalization for $d>1$, generate a  well defined Gaussian measure in a finite volume (remember $d < 4$).  In fact, the
determinant of  $D'_{{\rm local}}$ was calculated many times in the 
past~\cite{zinn}, and a generalization of a Quantum
Mechanical argument of ref.~\cite{au} indicates that this is all we need to compute the
determinant of $D'$.  The argument  goes as follows,
\begin{eqnarray}\label{det1}
{\rm Det} \left[  D'  \right] &=&  {\rm Det}  \left[ D'_{{\rm local}}  +   { 1 \over A}  |v> <v| 
\right]  \nonumber  \\
&=&  {\rm Det}  \left[ D'_{{\rm local}}   \right] 
\left(  1 +  { 1 \over A}    <v|  D^{' -1}_{{\rm local}}  |v>  \right).
\end{eqnarray}
Since $\varphi (u)$ is orthogonal  to  $\partial_{\mu} \varphi (u)$ (the zero modes
of  $D$ and $D_{{\rm local}}$), 
\begin{equation}\label{det2}
D'_{{\rm local}} \varphi =  D_{{\rm local}} \varphi = -2 \varphi^3.
\end{equation}
The last equality follows from the definition of  $D_{{\rm local}}$ in Eq.(\ref{Dlocal})
and the equation  (\ref{eq:numeqQM}) satisfied by  $\varphi$. Inverting 
$D'_{{\rm local}}$, and remembering  the definition  of $|v>$ and $A$ in Eqs.
(\ref{Dnonlocal}) and  (\ref{Adef}), we obtain
\begin{equation}\label{det3}
<v|  D^{' -1}_{{\rm local}}  |v>  =  -2 A.
\end{equation}
Replacing this result in Eq.(\ref{det1}), we arrive at the result
\begin{equation}\label{det4}
{\rm Det} \left[  D'  \right]  =  -  {\rm Det}  \left[ D'_{{\rm local}}   \right].
\end{equation}
As already mentioned, $D'_{{\rm local}}$ has one and only one negative
eigenvector, consequently its determinant is negative. Eq. (\ref{det4}) indicates
then that  ${\rm Det} \left[  D'  \right]$ is positive, as it should be according to 
(\ref{Dposdef}).  The effect of the nonlocal part is to change the sign of the 
determinant of the local part.

The preceding equations allow us to integrate the Gaussian approximation
of  $c_n   \left[ \varphi (u) \right]$ given in Eqs.~(\ref{local},\ref{nonlocal}).
Using the method of collective coordinates to project out the zero modes,
the Jacobian of the corresponding change of variables is, at leading order
in $1/ n$,
\begin{equation}\label{Jac}
J =  \prod_{\mu = 1}^{d}{  \left[ \int  \left( \partial_{\mu} \phi_{\rm max} \right)^2
{\rm d^d} x  \right]^{1/2}  }
\end{equation}
where no sum over $\mu$ is implied. 

It can be shown that the solutions of Eq.~(\ref{eq:eqQM}) corresponding to 
minimal action are spherically symmetric~\cite{zinn}, then (\ref{Jac}) can be
written as
\begin{equation}\label{Jac2}
J =   \left[  {1 \over {\rm d} } \int  \left( \partial_{\mu} \phi_{\rm max} \right)^2
{\rm d^d} x  \right]^{{\rm d} /2} 
\end{equation}
where now, sum over $\mu$ is implied. Using Eq.~(\ref{solusysteq1}) we
then find
\begin{equation}\label{Jac3}
J = n^{  {\rm d}  /2}.
\end{equation}
With this expression,  the functional integral of  $c_n   \left[ \varphi (u) \right]$
can be written as
\begin{eqnarray}\label{CalcFIcn1}
{1 \over  Z_0} \int  \left[ {\rm d} \phi  \right] \ c_n   \left[ \phi \right] &=& 
{ \left( -1 \right)^n  \over  2 \pi}  \left( n - 1 \right) !  \left( {\lambda \  m^{d-4}
\over A} \right)^n  \cdot  \\   \label{CalcFIcn2}
& & \left( {\rm Vol} \ m^d \right) n^{d/2}
\left(  -  {\rm Det}  \left[ {D'_{{\rm local}}  \over D_0} \right] \right)^{-1/2}
\end{eqnarray}
where$D_0 \equiv  - \nabla^2 + 1$. The factors 
in the line~(\ref{CalcFIcn1}) correspond to the value of  $c_n   \left[ \phi 
\right]$  at  $\phi_{\rm max}$ up to the normalization $1 / Z_0$ as can be
seen in Eq.~(\ref{cnphimax}). The factor ``Vol" arises after the integration
over  the flat coordinates corresponding to the center of $\phi_{\rm max}$.
The $n^{d/2}$ comes from the Jacobian of the change of variables as
mentioned before. The factor $m^d$ arises after the rescaling of the
fields that makes them dimensionless in both $c_n   \left[ \phi \right]$ and
$Z_0$. This happens because there are $d$ more integration variables in
$Z_0$ due to the integration over the collective coordinated in the numerator.
Finally, the factor 
$\left(  -  {\rm Det}  \left[ D'_{{\rm local}} \right] \right)^{-1/2}$ is the result
of the integration over the coordinates orthogonal to the zero modes of $D$,
while $\left( {\rm Det}  \left[ D_0\right] \right)^{1/2}$ is the dimensionless
normalization factor (the mass dimension of both, the numerator and the
denominator, was already taking care of in the term $m^d$). The minus sign is
due to the non-local part of $D$ that, as proved above, simply changes the
sign of the determinant of the local part, making it positive.

In the case $d=1$, 
$-  {\rm Det}  \left[ D'_{{\rm local}} /  D_0 \right] = 1/ 12$~\cite{zinn,au}, and Eqs.~(\ref{CalcFIcn1},\ref{CalcFIcn2}) (with $A= 4/3 $ as already
mentioned) become identical to the
corresponding result of Ref.~\cite{au} if we take into account the different 
normalization here and  a factor of 2 that is taken
care of  by remembering that the sign of the solution of  Eq.~(\ref{eq:eqQM})
is undetermined, therefore both, positive and negative solutions contribute
equally to the functional integral.

For $d = 2$ or 3, the formal expression~(\ref{CalcFIcn1},\ref{CalcFIcn2})
needs of course to be renormalized. All the arguments in this section remain
valid for the theory with a Pauli-Villars regularization~\cite{zinn}. The 
action~(\ref{eq:action}) becomes
\begin{equation} 
S  \left[ \phi \right] = \int   {\rm d^d} x  \left[ {1 \over 2}   \phi   \left( 
- \nabla^2 + {\nabla^4 \over \Lambda^2} + m^2 \right)  \phi 
+ {\lambda \over 4} \phi^4 + {1 \over 2} \delta m^2 (\Lambda) \ 
\phi^2 \right].
\label{eq:actionregul} 
\end{equation} 
The modification of the kinetic part of the action affects both the 
equation~(\ref{eq:eqQM}) and the scaling arguments, but by an amount
that becomes small like $\Lambda^{-2}$ when the ultra-violet cut-off
$\Lambda$ becomes large. 

As shown in Ref.~\cite{zinn}, although the counterterm increases with
the cut-off, since it is proportional to at least one power of $\lambda$,
taking the small $\lambda$ limit before the large cut-off limit  justifies
to ignore it in the equation~(\ref{eq:eqQM}) and the scaling arguments.
On the other hand it contributes to the result~(\ref{CalcFIcn1},\ref{CalcFIcn2}) 
an amount that exactly cancels the divergence in the
${\rm Det}  \left[ D'_{{\rm local}} \right]$ making the final expression finite
as it should be.

In the large $n$ limit, where the Gaussian 
approximation~(\ref{local},\ref{nonlocal}) becomes exact, the 
expression~(\ref{CalcFIcn1},\ref{CalcFIcn2}) gives the large order
behavior of the perturbative series of $Z$ (up to the factor of 2 mentioned
above) {\it without any assumption about the analytic structure in 
$\lambda$}~\cite{au}. A completely analogous procedure would give
the large order behavior of any Green's function.

Eqs.~(\ref{relphimaxvarphi}),~(\ref{cnphimax}),~(\ref{local},\ref{nonlocal})
and~(\ref{CalcFIcn1},\ref{CalcFIcn2}) , allow us to draw an accurate picture
of the mechanism underlying the lack of domination (in the sense of the
Lebesgue's theorem) of the convergence of the sequence of  perturbative
integrands~(\ref{eq:fNfunctint}) towards~(\ref{eq:exIntd}), and consequently 
of the mechanism underlying the divergence of the perturbative series. In
fact, perhaps not surprisingly given the similarity of their large order 
behavior, this picture is very similar to the one described in the previous
section for the simple integral example.

In a finite volume, there is a region of finite measure in field space in which
the perturbative approximation $f_N [ \phi (x) ]$ of  Eq.~(\ref{eq:fNfunctint})
approximate the exact integrand~(\ref{eq:exIntd}) with an error smaller
than a given prescribed number. This region grows with $N$, becoming the 
full field space in the $N \rightarrow \infty$ limit. As in the simple example, 
this is a consequence of  Taylor's theorem applied to the (analytic)
integrand~(\ref{eq:exIntd}).

The problem is that, for any finite $N$, outside that region the approximate 
integrand  $f_N [ \phi (x) ]$ strongly deviates from the exact one. This can be
seen by noting that the maxima of every term of $f_N$ grow factorially
with the order. Therefore, for large enough $N$, the last term is far greater
than the previous ones at its maxima. Even more, as shown above, the 
Gaussian approximation around that maxima (that becomes exact for 
$N \rightarrow \infty$) defines a measure that does not go to zero as 
$N \rightarrow \infty$ (in fact, it is independent of 
$N$~(\ref{local},\ref{nonlocal})). This means that for every finite $N$, 
there is a region of finite measure in field space (and this measure does not
go to zero as $N \rightarrow \infty$) in which the deviation between the 
perturbative integrand and the exact one is of the order of the maxima of the
last term of $f_N$, i.e., of the order of $\left( N - 1 \right)!$.  No integrable
functional can therefore satisfy the property  (\ref{eq:bound}) of the 
Lebesgue's theorem.

That is the mechanism that  makes the sequence of perturbative integrands,
although convergent to the exact one, non-dominated in the sense of
Lebesgue's theorem. That is therefore the mechanism that makes the sequence
of integrals (i.e., the perturbative series) divergent. In fact, as 
Eqs.~(\ref{CalcFIcn1},\ref{CalcFIcn2}) show,  the famous factorial 
behavior of the large order  coefficients of the perturbative series is a
consequence, after integration, of the above mechanism.

\section{Steps Towards a Convergent  Series}\label{sec:converg}

It was mentioned in the introduction that the analysis of the divergence
of perturbation theory presented in this paper would point directly towards
the aspects of the perturbative series that need to be modified in order
to generate a convergent series. This is the topic of the present section.

In the previous section we analyzed perturbation theory from the point of  view of 
the Dominated Convergence Theorem. We have detected the precise way in
which the convergence of the sequence of perturbative integrands to the exact
one takes place, and the way this convergence fails to be dominated.  We have
learned that for any finite order $N$, the field space naturally divides into two
regions. In the first one, that grows with the order, eventually becoming the full
field space (in the $N \rightarrow \infty$ limit), the perturbative integrands very
accurately approximate the exact one. In the other one, however, the deviation
between the perturbative and exact integrands is so strong, that the sequence
of integrals diverge.

 It is then clear that
{\it if we could somehow modify the integrands, order by order, in the region
where they deviate from the exact  one, while preserving them as they
are in the other region, then, with a ``proper" modification, such modified
sequence of integrands would converge in a dominated way. According to the
Dominated Convergence Theorem, their integrals would then converge to the exact
integral, achieving the desired goal of a  convergent modified perturbation
theory.}
\\

Let us call $\Omega_N$ to the region of field space in which the ${\rm N^{th}}$
perturbative integrand approximate with a given prescribed error the exact
integrand (\ref{eq:exIntd}). The {\it characteristic function}, 
${\rm Ch} (\Omega_N, \left\{ \phi (x) \right\})$, of that region is equal to 1 for
field configurations belonging to it, and zero otherwise:
\begin{equation}\label{chfunctome}
{\rm Ch} (\Omega_N, \left\{ \phi (x) \right\})  \equiv
\cases{
1 & for $ \left\{ \phi (x) \right\}  \in \Omega_N $ \cr
0 & for $ \left\{ \phi (x) \right\}  \not\in \Omega_N $}
\end{equation}

One possible realization of the above strategy of modifying the integrands
(\ref{eq:fNfunctint}) in the ``bad" region of field space is to make them zero
there. We would have
\begin{equation}\label{fprimach}
f_N' \left[ \phi (x) \right]=  {1 \over Z_0} \sum_{n=0}^{N}  
{\left( -1 \right)^n \over n!}   e^{- S_0  }
\left( {\lambda \over 4}   \int   {\rm d^d} x \phi^4   \right)^n  
{\rm Ch} (\Omega_N, \left\{ \phi (x) \right\})
\end{equation}

According to the analysis of the previous section, choosing $\Omega_N$
appropriately, the sequence of $f_N'  \left[ \phi (x) \right]$ will converge
dominatedly, and the corresponding interchange between sum and integral
will now be allowed. A rigorous proof of this is
left for a paper currently in preparation. For the purposes of the present
argument, it is sufficient to rely on the analysis of the previous section
to assume its validity. Also, in the next section we will analyze, along
the general ideas of this paper, some resummation schemes for which rigorous 
proofs of convergence have recently been given~[12-19]. As that
analysis will show,
these methods strongly rely on the general notions underlying the 
formula~(\ref{fprimach}). Their convergence supports, then,  the validity of the
dominated nature of the convergence of~(\ref{fprimach})
towards~(\ref{eq:exIntd}).

An urgent issue, however, is the practical applicability of the above
strategy. To implement it, we need a functional representation of the
characteristic function~(\ref{chfunctome}) (or an approximation to it) that 
only involves
{\it Gaussian and polynomial} functionals. In the same way in which a 
functional representation of the Dirac delta function allow us to perform functional integrals with constraints, the Fadeev-Popov quantization of Gauge
theories being the most famous example, a functional representation of the 
characteristic function~(\ref{chfunctome}) would allow us to functionally
 integrate only the
desired region of functional space.  Since, basically, the functionals we know
how to integrate reduce to Gaussians multiplied by polynomials, the desired
representation of the characteristic function should {\it only} involve those
functionals. Conversely, if it only involves those functionals, all the
sophisticated machinery developed for perturbation theory (including all
the perturbative renormalization methods) would automatically be applicable. 
With this in mind, consider the following function,
\begin{equation}\label{win}
W (M, u) \equiv e^{-Mu} \sum_{j=0}^{M} {\left( Mu \right)^j \over j! }
\end{equation}
where $M$ is a positive integer.  Note that $W (M,u)$ arises from 
$1 = e^{-Mu} e^{+Mu}$ by expanding the second exponential up to order M.
$W (M, u)$ has the following remarkable properties,
\begin{enumerate}
\item  $W(M,u) \rightarrow 1 $  when $M \rightarrow  \infty$ for $0<u < 1$. 
           The convergence is uniform, with the error going to zero as
\begin{equation}\label{item1}
R (M,u) \le e^{M  \left(  \ln{u} - (u - 1) \right)} {1 \over  \sqrt{2 \pi M}}
{u \over 1 - u + 1/M}.
\end{equation}

\item  $W(M,u) \rightarrow 0 $  when $M \rightarrow  \infty$ for $1<u $. The 
           convergence is also uniform , with an error of the form
\begin{equation}\label{item2}
W (M, u) \le e^{M \left(  \ln{u} - (u - 1) \right)}.
\end{equation}
\end{enumerate}

As we see, the exponent corresponds to the same function in both cases. 
For $u > 0$, this function is always negative except at its maxima, at $u =1$,
where it is 0. Therefore the convergence is in both cases exponentially fast
in $M$, with the exponent becoming more and more negative, for a fixed $M$,
when  $u$ differs more and more from 1. The proof  of  properties 1 and 2 is in 
the Appendix 1.

If we replace $u$ by a positive definite quadratic form $< \phi | D | \phi
>/C_N$, then the insertion of Eq.~(\ref{win}) into the functional integral would
effectively cut off the  region of integration $< \phi | D | \phi > \ > C_N$
\begin{eqnarray}\label{fprimaw}
Z_N' \left[ \phi (x) \right] &=&  {1 \over Z_0} \int   \left[ {\rm d}  \phi \right]
\sum_{n=0}^{N}  {\left( -S_{\rm Int} \right)^n \over n!}   e^{- S_0  } 
  \lim_{M \rightarrow \infty} W \left( M, {< \phi | D | \phi> \over  C_N}
 \right)    \\
&=& {1 \over Z_0} \sum_{n=0}^{N}  {\left( -1 \right)^n \over n!} 
\lim_{M \rightarrow \infty}  \int  \left[  {\rm d}  \phi \right]   e^{- S_0  } 
\left( S_{\rm Int} \right)^n W 
\left( M, {< \phi | D | \phi> \over  C_N} \right) \nonumber  \\ \label{fprimaw2}
\end{eqnarray}
$C_N$ is a constant that changes with the order  $N$ of the expansion in 
$\lambda$, increasing with $N$ but in such a way that in the region 
$< \phi | D | \phi>  < C_N$ the difference between the perturbative  and the 
exact integrands is smaller than a given prescribed error. Since the
convergence of $W$ is uniform according to properties
1 and 2, with errors given in Eqs.~(\ref{item1}) and~(\ref{item2}), the
corresponding interchange between the sum in Eq.~(\ref{fprimaw2}) 
and the functional
integral is justified. The fact that $u$ becomes a {\it quadratic} form implies that 
the resulting integrands are Gaussians multiplied by monomials, therefore the
familiar Feynman diagram techniques can be used to integrate them. It
also implies that no new loops appear and the sum in $j$
from~(\ref{win}) becomes an algebraic problem. A typical functional integral 
to compute has the form
\begin{equation}\label{typicalfunctint}
\int   \left[ {\rm d}  \phi \right] e^{- \int   {\rm d^d} x  \left[ {1 \over 2}  
\left( \partial_{\mu} \phi \right)^2 + {1 \over 2}  m^2 \phi^2 
  + ( \phi  D  \phi  / C_N )    \right] }  \left( \int {\rm d^d} x \phi^4 \right)^n
\left( \int {\rm d^d} x  \ \phi D  \phi \right)^m
\end{equation}
as can be seen by replacing the definition~(\ref{win}) into~(\ref{fprimaw2})
with $u = < \phi | D | \phi > / C_N$. 

Note that at any given order in $\lambda$, it is not necessary in principle
to go to infinity in $M$. That would amount to replace the 
perturbative integrands by zero in the region $< \phi | D | \phi > \ > C_N$,
realizing the strategy mentioned before. But since the convergence in $W$
is uniform, a finite, large enough $M$ (depending on the order in the expansion
in the coupling constant), would suffice to tame the behavior
of the perturbative integrands and transform them into a {\it dominated}
convergent sequence. In fact, as we will see, many methods of improvement
of perturbation theory use effectively formula~(\ref{win}) without sending
$M \rightarrow  \infty$ for any given finite order in perturbation theory.
In any case, as already mentioned, that limit is in principle computable, since
it does not involves new loops. Work in this direction is in progress.

The convergence of the sequence~(\ref{fprimaw2}) towards $Z (\lambda)$ 
may be thought, at first sight, to be in conflict with our well established
knowledge about the non-analyticity of this function at $\lambda = 0$. In fact,
Eq.~(\ref{fprimaw2}) seems to be a power series in $\lambda$ (the powers of
$\lambda$ coming from the powers of $S_{\rm Int}$), therefore, if convergent,
that power series would define a function of $\lambda$ analytic at 
$\lambda = 0$. It must be recognized, however, that the
validity of Lebesgue's Dominated Convergence Theorem is completely
independent of any analyticity consideration. Therefore, if  its hypothesis are
satisfied, its conclusions must be valid. This being said, the question of how
does the convergence of~(\ref{fprimaw2}) fits with the non-anlayticity of $Z (
\lambda )$ deserves an answer. To begin with, even at finite order in $\lambda$,
the function~(\ref{fprimaw2}) is not necessarily analytic at $\lambda =0$ despite
its analytic appearance. This is because the constant $C_N$ may have an
implicit nonanalytic dependence on $\lambda$. In Appendix 2 this is actually
the case in the context of a simple example to which the present ideas are
applied. But the mechanism that ultimately introduces the proper non-analyticity
in $\lambda$ is the limit process $N \rightarrow \infty$. Given a non-analytic
function like $Z (\lambda)$ one can always construct a sequence of analytic
functions that converge to it. Satisfying the hypothesis of the Dominated
Convergence Theorem  is a way of achieving that, avoiding all the
complicated and {\it model dependent} issues of non-analyticity. Note that the
validity of these hypothesis for a given sequence of integrands can be
checked independently of any analyticity consideration.

\begin{figure}[htp]
\hbox to \hsize{\hss\psfig{figure=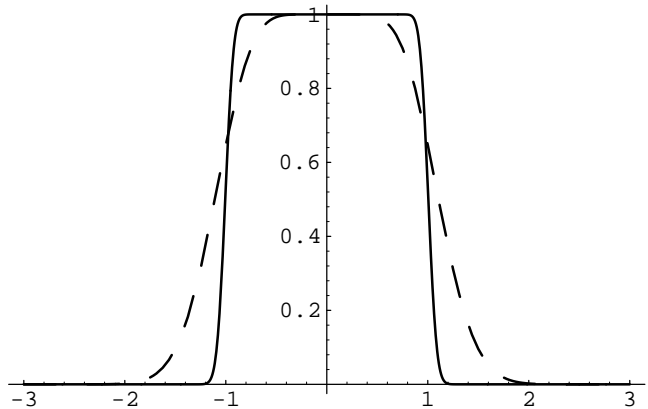,width=0.7\hsize}\hss}
\caption{Function $W(M,x,x_{c,N})$ with $x_{c,N}=1$ for $M=3$ (segmented
line) and $M=60$ (continuous line). The convergence towards the 
characteristic function of the interval $|x|< x_{c,N}$ is apparent.}   
\label{fig3}
\end{figure}

In the Appendix 2 we prove the convergence of  the general strategy 
discussed here for the simple integral example analyzed in
section~(\ref{sec:simpleint}). For that  case, making $u = (x/x_{c,N})^2$, the
function $W (M, u)$ becomes in the limit the characteristic function of the 
interval $|x|< x_{c,N}$. We use this to explicitly compute the non-analytic
function $z (\lambda)$ (Eq.~(\ref{eq:pertsimple})) calculating only Gaussian
integrals. We also show  explicitly how a non-analytic dependence of  
$x_{c,N}$ on $\lambda$ naturally arises just by demanding  the validity
of the Lebesgue's hypothesis and how the $N \rightarrow \infty$ limit
process captures the full non-analyticity of $z (\lambda)$. The same 
method also works for the ``negative mass case", where the Borel
resummation method fails. In figure~\ref{fig3}, we can appreciate the
convergence of $W$ towards the characteristic function of  the interval
$|x|< x_{c,N}$ for $x_{c,N} = 1$ for two different values of $M$.

\section{Improvement methods of perturbation theory.}\label{Improvement}

The analysis of the mechanism of  divergence of the perturbative series
presented in this paper, together with the formula~(\ref{win}) and its
properties, offer a large range of possibilities to construct a convergent
series. In the previous section we have shown how that formula can be 
used to effectively cut-off  the region of field space where the strong
deviation between perturbative and exact integrands take place. But as we will
see, this is only one possible way, among many, to use the formula~(\ref{win})
to transform the sequence of perturbative integrands into a dominated one.

Another example of its possible use is the so called ``optimized delta
expansion"~\cite{gral, stev}. In a series of papers~\cite{tony1}-\cite{kle},
it was proved that  such an expansion converges for the partition function 
of the anharmonic oscillator in finite Euclidean time. The problem of 
convergence in the infinite Euclidean time (or zero temperature) limit for the
free energy or any connected Green's function is still under investigation,
as well as its extension to Quantum Field Theories~\cite{tony3,arv}. 
The method was proved to generate a convergent series for the energy 
eigenvalues~\cite{gui1, gui2}, although such studies make heavy use of
analyticity properties valid specifically in the models studied. In these 
works, it was realized that many methods of improvement of perturbations 
theory, such as the order dependent mappings of references~\cite{zinn2, leg},
posses the same general structure as the linear delta expansion. 
A considerable amount of work has been dedicated  to investigate the virtues
and limitations of the method and extensions of it~\cite{kle, nev}.

It is not the place here to give a detailed analysis of these methods. But we
would like to briefly indicate how they can be 
understood in terms of the ideas presented here. In what follows, our analysis
is restricted to $d=1$ (Quantum Mechanics) where rigorous results
about the convergence of the methods considered here are available.

Let's consider the case of the anharmonic oscillator. Its action is given
in Eq.~(\ref{eq:action}) for $d=1$. The idea of the method is to replace it
by an interpolating action
\begin{equation}\label{sdelta}
S_{\delta} = \int   {\rm d} t  \left[ {1 \over 2}  \left( {\rm d}_t \phi \right)^2
+ {1 \over 2}  \left( m^2 + {\lambda \over 2 m} \alpha \right) \phi^2 + 
\delta {\lambda \over 4} \left( \phi^4  -{ \alpha \over m} \phi^2 \right) \right].
\end{equation}
Clearly, the dependence on the parameter $\alpha$ in $S_{\delta}$ is lost
when $\delta = 1$. For that value, the action~(\ref{sdelta}) reduces 
to~(\ref{eq:action}). However, if we expand up to a finite order in $\delta$
and then make $\delta=1$, the result still depends on $\alpha$. The idea is
to tune $\alpha$, order by order in the expansion in $\delta$, so that the result
is a convergent series. It was shown in the references mentioned above that
the methods works if $\alpha$ is tuned properly. For example, in 
reference~\cite{tony2}, the asymptotic scaling $\alpha \simeq N^{2/3}$
was used to prove the convergence of the method for the partition
function at finite Euclidean time.

It is interesting to note that originally~\cite{tony1, tony2}, $\alpha$ was
tuned according to heuristic prescriptions such as the ``principle of minimal 
sensitivity"~\cite{stev} (at any given order in $\delta$, choose $\alpha$ so
that the result is insensitive to small changes in it), or the criterion of
``fastest apparent convergence" (the value of $\alpha$ at which the next order
in delta vanishes). But later~\cite{tony3, arv},
it was realized that the best strategy was simply to leave $\alpha$
undetermined, find an expression for the error (that obviously depends on
$\alpha$), and then choose $\alpha$ so that the error goes to zero when
the order in $\delta$ goes to infinity.  It is clear that a {\it structural}
understanding of the convergence of the method can help to construct the
necessary generalizations to overcome the difficulties associated with the
convergence in the infinite volume limit for connected Green's functions, as
well as the extensions to general Quantum Field Theories. 

To understand the ``optimized delta expansion" in terms of the ideas presented
in this paper, let us expand the functional integral corresponding to the
action~(\ref{sdelta}) in powers of $\delta$ up to a finite order $N$, and make
$\delta =1$ as the method indicates,
\begin{eqnarray}\label{delta1}
Z (m, \lambda, \alpha, N) &=& {1 \over  Z_0} \int  \left[ {\rm d} \phi  \right]
e^{-  \int   {\rm d} t  \left[ {1 \over 2}  \left( {\rm d}_t \phi \right)^2
+ {1 \over 2}  \left( m^2 + {\lambda \alpha \over 2 m}  \right) \phi^2  \right] } 
\cdot   \nonumber  \\
& & \left[  \sum_{n=0}^N  {\left( - 1 \right)^n \over n!}  \left( 
 {\lambda \over 4} \int \phi^4  - {\lambda   \alpha \over 4 m}
 \int \phi^2 \right)^n  \right]
\end{eqnarray}

The general analysis of the mechanism of divergence of perturbation theory
of section~\ref{sec:leb} indicates that if  the function~(\ref{delta1}) generates
a convergent series with $\alpha$ scaling properly with $N$, then, barring
miraculous coincidences, the corresponding integrands should converge
dominatedly (or, even better, uniformly) towards the exact
integrand~(\ref{eq:fNfunctint}). We want to obtain a qualitative understanding 
on how this method achieves that.

Expanding the binomial and making some elementary changes of variables in the 
indices of summation, we obtain the expression
\begin{eqnarray}\label{delta2}
Z (m, \lambda, \alpha, N) &=& {1 \over  Z_0} \int  \left[ {\rm d} \phi  \right]
e^{-  \int   {\rm d} t  \left[ {1 \over 2}  \left( {\rm d}_t \phi \right)^2
+ {1 \over 2}  \left( m^2 + {\lambda \alpha \over 2 m}  \right) \phi^2  \right] } 
\cdot   \nonumber  \\
& & \left[  \sum_{i=0}^N  {\left( - 1 \right)^i \over i!}  \left( 
 {\lambda \over 4} \int \phi^4 \right)^i  
\left( \sum_{k=0}^{N-i} {1 \over k!}  \left( {\lambda   \alpha \over 4 m} 
\int \phi^2 \right)^k \right) \right]
\end{eqnarray}
This equation already shows some of the distinctive characteristics of
the method. As we see, the ${\rm i^{th}}$ power of the interacting action in the
expansion of $e^{-S_{\rm Int}}$ up to order $N$, is multiplied by
\begin{equation}\label{delta3}
{\cal W} \left( N - i \right) \equiv  e^{- \left( \lambda \alpha  / 4 m \right)
 \int \phi^2}  \left( \sum_{k=0}^{N-i} {1 \over k!}  \left( {\lambda   \alpha \over 4 m}
 \int \phi^2 \right)^k \right).
\end{equation}
Note that ${\cal W} \left( N  \right)$ corresponds to the function $W (M, u)$
with $M = N$ ($N$ is the order in the expansion of  $e^{-S_{\rm Int}}$), and
the variable $u$  replaced by the quadratic form $\left( \left( \lambda  / 4 m
\right) \int \phi^2 \right) / C_N$, where  $C_N = N / \alpha$. Making for example
$\alpha \simeq N^{2/3}$ as in Ref.~\cite{tony2} (where it was proved that 
with such scaling
the method generates a convergent series), we see then, that, according to the
previous section, ${\cal W} \left( N  \right)$ is an approximation of the theta
function in the region of field space characterized by
\begin{equation}\label{delta4}
{\lambda   \alpha \over 4 m} \int  {\rm d} x  \ \phi^2 \le  N^{1/3}.
\end{equation}

Equation~(\ref{delta2}), however,  shows that the mechanism used to 
achieve dominated convergence can not be reduced to a simple insertion of 
the function $W(M,u)$ with $M=N$ and 
$u = \left( \left( \lambda  / 4 m \right) \int \phi^2 \right) / C_N$. 
That would be the case if all the powers of the expansion of  $e^{-S_{\rm Int}}$
up to order $N$ were multiplied by ${\cal W} \left( N  \right)$. But
equation~(\ref{delta2}) shows that the ${\rm i^{th}}$ power of the interacting
action is in fact multiplied by ${\cal W} \left( N - i  \right)$. 
\\

At this point it is convenient to pause for a moment in our study of the 
``optimized delta expansion" to give some useful definitions.

Let us call {\bf passive} mechanisms  (to achieve dominated, or uniform
convergence of a sequence of integrands to the exact one) to those that
can be reduced to the product of the $N^{\rm th}$ perturbative integrand
and the characteristic function of  a region $\Omega_N$ of field space for
some sequence  $\left\{ \Omega_N \right\}$.

Passive methods use only  information that is already available in the
perturbative integrands, they just get rid of the ``noise" inherent to perturbation
theory. Because of that, in addition to define a convergent series,  they can 
also be very useful to study perturbation theory itself. The function $W (N,u)$,
with $u$ replaced by a properly selected quadratic operator, was specially
designed to make passive methods practical. In a sense, section~\ref{sec:converg} is a discussion  of passive methods.

{\bf Active} mechanisms are those that are not passive, as defined above.
\\

What kind of mechanism is the one underlying the ``optimized delta expansion"
method?

A trivial generalization of the proof, in the previous section, of the convergence
of $W(M,u)$ towards the theta function for $u>0$, shows that the function
\begin{equation}\label{wbarra}
\overline{W} (M,u,i) \equiv e^{-Mu}  \sum_{n=0}^{M-i} 
{\left( M u \right)^i  \over i!}
\end{equation}
also converges towards the theta function for $u>0$ in the limit 
\begin{equation}\label{limit}
M \rightarrow \infty, \quad  i \ {\rm fixed}.
\end{equation}
 In this sense, the ``optimized delta expansion" method
does have passive aspects. As Eqs.~(\ref{delta2}, \ref{delta3}) show, it
amounts to multiplying the ${\rm i^{th}}$ power of  the expansion up to 
order $N$ of $e^{S_{\rm Int}}$ by $\overline{W}$ with $u = \left( \left( 
\lambda  / 4 m \right) \int \phi^2 \right) / C_N$ and  $C_N = N / \alpha$. 
Since this function converges to the characteristic function of the region
characterized by Eq.~(\ref{delta4}), this means that the first $i$ terms, of
the expansion up to order $N$ of $e^{S_{\rm Int}}$ are {\it effectively}
multiplied by the same function (an approximate characteristic function) for  
$i \ll N$ . Therefore, the first $i$ terms, with $i \ll N$,  use only the
information available in the perturbative series to converge to the exact
integrand. 

What about the other terms?, the ones characterized by $i {\ \lower-1.2pt\vbox{\hbox{\rlap{$<$}\lower5pt\vbox{\hbox{$\sim$}}}}\ } N$?
Surprisingly, these terms produce a convergence of the corresponding
integrands towards the exact one that is faster than possible with only passive
components!

It is not the place here to study this aspect in detail, so, let us simply see
this ``faster than passive" convergence for the simple integral example.

Applied to the ``mass-less" version of the integral~(\ref{eq:pertsimple}), the
optimized delta expansion method was proved to generate a rapidly convergent
sequence in Ref.~\cite{tony1}. That is, the sequence given by
\begin{equation}\label{simODE}
I_N \equiv \sum_{n=0}^N {(-1)^n \over n!} \int_{- \infty}^{\infty} {\rm d}  x  \
e^{ - \alpha (N) \ x^2}   ({\lambda \over 4} x^4- \alpha (N) \  x^2)^n 
\end{equation}
was proved to converge to 
\begin{equation}\label{sim}
I \equiv \int_{- \infty}^{\infty} {\rm d}  x   \ e^{ -  \lambda  x^4 / 4} 
\end{equation}
when $\alpha (N) \simeq \sqrt{N}$ with an error that goes to zero at the 
very fast rate of  $R_N  < C N^{1/4} e^{-0.663 N}$ when $N \rightarrow \infty$.
$C$ is a numerical constant.

\begin{figure}[htp]
\hbox to \hsize{\hss\psfig{figure=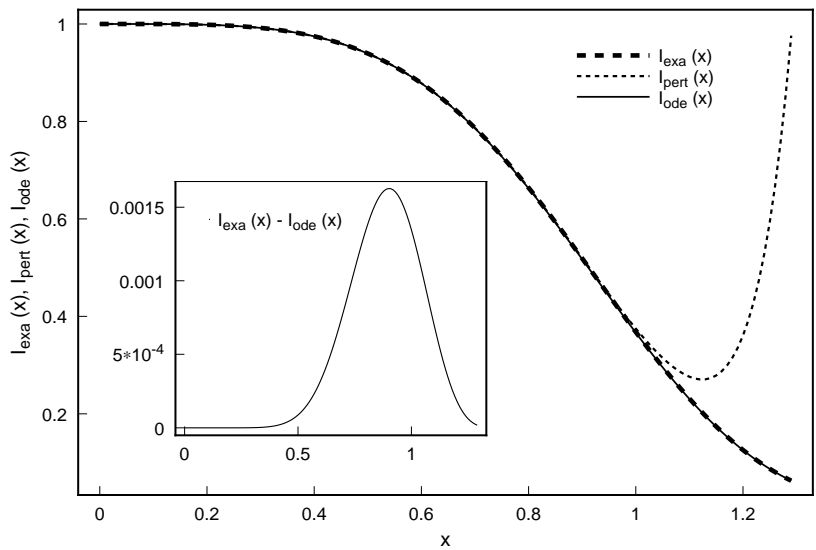,width=0.9\hsize}\hss}
\caption{In the main plot, the superiority of the convergence of the fourth
order optimized delta expansion (ode) with respect to the same order 
perturbative approximation is evident. In the sub-graph, the difference 
between the ode and the exact integrand is plotted. Note the difference in
the scales of the y axis of  the main and sub graph.}   
\label{fig4}
\end{figure}

We are interested in understanding whether the corresponding convergence
of the integrands is faster than passive. 
For our qualitative purposes, it is
enough to observe, in figure~\ref{fig4}, the convergence towards the exact
integrand
\begin{equation}\label{exinm0}
I_{\rm exa} (x)  =   e^{ -  {\lambda \over 4}  x^4}
\end{equation}
of both, the perturbative integrand
\begin{equation}\label{pertintm0}
I_{\rm pert} = \sum_{n=0}^N { ( - \lambda   x^4 / 4)^n \over n!}   
\end{equation}
and the optimized delta expansion integrand
\begin{equation}\label{odeinm0}
I_{\rm ode}  = \sum_{n=0}^N {(-1)^n \over n!} 
e^{ - \alpha (N) \ x^2}   ({\lambda \over 4} x^4- \alpha (N) \  x^2)^n
\end{equation}
with $\alpha (N) \simeq  \sqrt{N}$, for $N=4$.

We can see how accurate the convergence of  $I_{\rm ode} (x)$ is, already
at this low order. In particular, when the perturbative integrand begins to
diverge, $I_{\rm ode} (x)$ continues to approximate remarkably well  the
exact integrand. In the sub-figure, we can appreciate the difference
between $I_{\rm exa} (x)$ and $I_{\rm ode} (x)$. Note the difference in the
$y$ axis scale of graph and sub-graph.

It is then clear that the optimized delta expansion method, with its subtle
combination of passive and active components, manages to generate
a sequence of integrands that  (uniformly) converges towards the exact
one at a rate that far exceeds the possibilities within a purely passive method.

We stop here the qualitative discussion of the optimized delta expansion
method expressing two general lessons:
\begin{enumerate}
\item Any method of improvement of the perturbative series in a given
quantum theory, where a functional integral representation of the quantity
under study exists, must rely, at the level of the integrands,  on an
improvement over the pointwise convergence of the Taylor series in the
coupling constants of $e^{-S}$.
\item The problem of finding a convergent series reduces to the problem
of finding a dominatedly convergent sequence of  integrands towards $e^{-S}$. 
\end{enumerate} 
This second simple statement, not only provides a guide to the construction
of convergent schemes, but also emphasizes the fact that, in principle, a
dominatedly convergent sequence of integrands, do not have to have
any relation whatsoever with the corresponding Taylor expansion. In order
to be able to use the usual techniques of  Quantum Field Theory, it is
reasonable to restrict the search for a convergent scheme to a sequence of
integrands of the general form:
\begin{equation}\label{generic}
f_N = e^{-S_0} \sum_{n=0}^N a_n \ S_{\rm Int}^n  \ f_n (< \phi | D_n | \phi >)
\end{equation}
where the functional $f_n$ of the quadratic form $< \phi | D_n | \phi >$, 
should take care of the non-dominated convergence that is bound to
appear with only powers of the interacting action. The function $W$ of
section~\ref{sec:converg}, with its possible generalizations, is an ideal
candidate for this purpose. But the selection of the coefficients $a_n$
amounts to a pure problem in optimization of the convergence of the
integrands $-$ no a priori connection with any Taylor series is necessary.

\section{Conclusions}\label{sec:concl}

In this paper we have exposed the mechanism, at the level of the integrands,
that make the perturbative expansion of a functional integral divergent.
We have seen in detail how the sequence of integrands violate the domination
hypothesis of the Lebesgue's Dominated Convergent Theorem. That theorem,
as is well known, establishes the conditions under which it is allowed
to interchange an integration and a limit, in particular the one that takes
place in the generation of perturbation series.

It was shown that at any finite order in perturbation theory, the field
space divides into two regions. One, that grows with the order, in which
the perturbative integrands very accurately approximate the exact integrand.
In the other however, a strong deviation takes place. It was shown that the
behavior in this second region violates the hypothesis of Lebesgue's
theorem, and, consequently, generates the divergence of perturbation theory.
The famous factorial growth of the large order coefficients of the perturbative
series was shown to be an effect, after integration, of the very mechanism that
violates the hypothesis of the theorem.

All of the above was done explicitly without relying in the particular
analytic properties of the models studied. It is therefore natural to 
assume that similar mechanisms of violation of the Lebesgue's hypothesis
are present in any other Quantum Field Theory, although for just
renormalizable theories other mechanisms are responsible for renormalons.
Studies in this direction are in progress.

The mechanism of divergence presented here points towards a simple way
to achieve a convergent series: integrate only in the ``good" region of 
field space. Since this region grows with the order, becoming in the limit
the whole field space, integrating in a correspondingly increasing region
we would obtain a convergent series. A step forward towards a practical
implementation of this program was made with the construction of the 
function $W$~(\ref{win}). This function allows us to introduce a Gaussian 
representation of the characteristic function of regions of field space,
very much like the imposition of constraints in the functional integral
was allowed by a functional representation of the Dirac's delta function.
A rigorous proof of the convergence of this practical implementation of the
above  mentioned strategy is in progress. In Appendix 2 it was applied to
a simple integral example.

Finally, a qualitative analysis of the optimized delta expansion method
of improvement of perturbation theory in terms of the ideas of this paper
was done. Some general properties of improvement methods, useful to
generates new schemes, as well as to understand and improve old ones,
have been established.

\newpage

\section*{Acknowledgments}

S. P.  was supported in part by U.S. Dept. of Energy Grant  DE-FG
02-91ER40685. He would like to thank  A. Duncan for numerous very
useful discussions, and also to A. Das and S. Rajeev.

\section*{Appendix 1}

In this appendix we will prove the two properties of formula~(\ref{win}).

Because for $u>0$ all the terms of the sum defining $W (M, u)$ are positive,
we have trivially $W (M, u) > 0$. On the other hand, since in the Taylor 
expansion of $e^{Mu}$ all the terms are positive, we have
$\sum_{n=0}^{M} \left( Mu \right)^n / n!   \le e^{Mu}$.
Therefore $W (M, u) \le 1$. So for every $M$ and positive or zero $u$ we
have,
\begin{equation}\label{app:bound}
0 \le W(M,u) \le 1.
\end{equation}

Consider first the case $0<u < 1$.
\begin{equation}\label{app:rest}
1 - W (M, u) = e^{-Mu} \sum_{n=M+1}^{\infty} {\left( Mu \right)^n \over n! }
\equiv  R(M,u),
\end{equation}
we will prove that  $R(M,u) \rightarrow 0$ when $M \rightarrow \infty$.

Changing variables to $j = n -M$, we get
\begin{eqnarray}\label{app:rest2}
 R(M,u) &=& e^{-Mu} {\left( Mu \right)^M \over M! } \sum_{j=1}^{\infty} 
\left( Mu \right)^j {M! \over \left( j+M \right)! }   \\
&\le& e^{-Mu} {\left( Mu \right)^M \over M! } \sum_{j=1}^{\infty} 
{\left( Mu \right)^j  \over \left( M + 1 \right)^j }  \\
&\le& e^{-Mu} {\left( Mu \right)^M \over M! }  {u \over 1 - u + 1/M}.
\end{eqnarray}
But $M^M / M! \rightarrow  e^M /  \sqrt{2 \pi M}$ for large $M$, so
\begin{equation}\label{app:rest3}
R (M,u) \le e^{M  \left(  \ln{u} - (u - 1) \right)} {1 \over  \sqrt{2 \pi M}}
{u \over 1 - u + 1/M}.
\end{equation}
The exponent is negative in the region $0<u < 1$ since, being both,
$\ln{u}$ and $(u - 1)$ negative there, $| \ln{u} | > |u - 1|$ in this region. 
Therefore
\begin{equation}\label{app:proved}
R (M,u) \rightarrow 0 \ , {\rm when} \quad  M \rightarrow \infty
\end{equation}
in the region $0<u < 1$ and property 1 is proved with an exponentially fast
convergence.

In the region $u > 1$, we have
\begin{eqnarray}\label{app:uge1}
W (M, u) &=& e^{-Mu} \sum_{n=0}^{M} {\left( Mu \right)^n \over n! } \le
e^{-Mu} u^M \sum_{n=0}^{M} {M^n \over n! }   \\  \label{app:uge2}
&\le& e^{M  \left(  \ln{u} - (u - 1) \right)}.
\end{eqnarray}
The first inequality is valid because $u>1$ and the second because  \\
$e^M > \sum_{n=0}^{M} M^n / n!  $. The exponent is again negative.
For $u>1$, both,  $ \ln{u}$  and   $( u-1)$ are positive, but now 
$| \ln{u} | < |u - 1|$.  So property 2 is also valid with an exponentially fast
convergence. 

For $u=1$ all we know is that $W$ is bounded by Eq.(\ref{app:bound}). That is
all we need. Numerics suggest $W (M, 1) \rightarrow 1/2$ when $M \rightarrow
\infty$.

This finishes our proof.

\section*{Appendix 2} 

In this appendix we apply the strategy discussed in section~\ref{sec:converg}
to generate a series convergent to the function $z (\lambda)$ 
(Eq.~(\ref{eq:pertsimple})). This is done using the function $W$ of 
Eq.~(\ref{win}) and computing {\it exclusively} Gaussian integrals, therefore
we restrict ourselves  to using only those techniques that are also available in
Quantum Field Theory.

As mentioned in section~\ref{sec:converg}, the simplest possible modification
of the perturbative integrand (\ref{eq:fNsimpltint}) that would transform the
corresponding sequence into a dominated one, amounts to keep them as they
are for $|x| < x_{c,N}$ and replacing them by zero for $|x| > x_{c,N}$.  That is,
\begin{equation}\label{convseq}
f'_N  = \cases{  \pi^{-1/2}  \sum_{n=0}^{N}  {\left( -1 \right)^n \over n! } 
\left( {\lambda \over 4} x^4 \right)^n    e^{- x^2   } & for $|x| < x_{c,N} $ \cr
0 & for $|x| > x_{c,N}$}.
\end{equation}
In fact, choosing $x_{c,N}$ so as to properly avoid the region in which the 
deviation takes place, the sequence of $f'_N$ converges {\it uniformly} towards
the exact integrand (\ref{eq:exintd}) as we will show shortly. Consequently, the
corresponding  sequence of integrals
\begin{eqnarray}\label{exint2}
\int_{- \infty}^{\infty} {\rm d} x   f'_N  & = & \pi^{-1/2}  \int_{- x_{c,N} }^{x_{c,N}}  {\rm d} x 
 \sum_{n=0}^{N}    {\left( -1 \right)^n \over n! } 
\left( {\lambda \over 4} x^4 \right)^n    e^{- x^2   }  \\  \label{exint3}
& = & \pi^{-1/2}  \sum_{n=0}^{N}  \int_{- x_{c,N} }^{x_{c,N}} {\rm d} x  
{\left( -1 \right)^n \over n! }  \left( {\lambda \over 4} x^4 \right)^n    e^{- x^2   }
\end{eqnarray}
will converge to the desired integral
\begin{equation}\label{desint}
z (\lambda) = \pi^{-1/2} \int_{- \infty}^{\infty} {\rm d} x    
e^{ - \left(  x^2 +  {\lambda \over 4} x^4 \right)  }.
\end{equation}
 In (\ref{exint2}) the change in the limits of  integration from $\pm \infty$ to 
$\pm x_{c,N}$ is just due to the definition of  $f'_N$ in Eq.(\ref{convseq}).
The interchange between sum and integral in (\ref{exint3})  is  now allowed 
because in the region $\left[  - x_{c,N}, x_{c,N}  \right]$ we have uniform 
convergence (this is a stronger condition than dominated convergence). The 
resulting integrals are not Gaussian due to the finite limits of integration. We will
show how they can be calculated using only Gaussian integrals.

A trivial way to achieve convergence of the sequence of integrals of the
$f'_N$ of Eq.~(\ref{convseq}) towards~(\ref{desint}) amounts to keep 
$x_{c,N}$ equal to a finite constant $``a"$ independent of N, while taking the limit
$N \rightarrow \infty$. In this limit, Eq.~(\ref{exint3}) becomes identical to
$\pi^{-1/2} \int_{-a}^{a} {\rm d} x e^{ - \left(  x^2 +  {\lambda \over 4} x^4 \right)  }$,
since for finite $a$ the Taylor series of the integrands converge uniformly.
Therefore, as already said, the interchange between sum and integral
is legal. Finally, taking the limit $a \rightarrow \infty$, we would obtain the
desired convergence towards $z (\lambda)$.

However, better use can be made of the information available in $f'_N$ for finite
$N$. For example, for every finite $N$, we can choose $x_{c,N}$ so that
\begin{equation}\label{diff}
|f'_N (x) - f (x) | \le {\epsilon_{T,N} \over 2 x_{c,N}} \quad {\rm for} \quad
|x| < x_{c,N},
\end{equation}
with $\epsilon_{T,N}$ 
going to zero as $N \rightarrow \infty$. Then, since  we have
\begin{equation}\label{diff2}
|f'_N (x) - f (x) | \le e^{- \left( x_{c,N}^2 + {\lambda \over 4} x_{c,N}^4 \right)}
\equiv  {\epsilon_{c,N} \over 2}  \quad  {\rm for} \quad |x| > x_{c,N},
\end{equation}
the $f'_N (x)$ will uniformly converge towards the exact integrand $f(x)$ 
if~(\ref{diff}) is consistent with $x_{c,N} \rightarrow \infty$ when 
$N \rightarrow \infty$. Indeed, if this happens, we would have
\begin{equation}
| \int_{- \infty}^{\infty}  \left( f(x) - f_N (x)  \right) {\rm d} x |   \le 
 \epsilon_{T,N} + \epsilon_{c,N} \rightarrow 0\quad {\rm when} \quad
N \rightarrow \infty.
 \end{equation}
The term $\epsilon_{T,N}$ comes trivially from~(\ref{diff}), while $\epsilon_{c,N}$
comes from~Eq.(\ref{diff2}) and the inequality 
\begin{equation}\label{esterr}
\int_{ x_{c,N} }^{\infty}  e^{- \left( x^2 + {\lambda \over 4} x^4 \right)} {\rm d} x
\le e^{- \left( x_{c,N}^2 + {\lambda \over 4} x_{c,N}^4 \right)} = \epsilon_{c,N},
\end{equation}
valid for $x_{c,N}>1$.

Applying Taylor's theorem to the function $e^{- \lambda x^4 / 4}$ one can
easily show that the condition~(\ref{diff})  is satisfied if 
\begin{equation}\label{xcNasfunctN}
x_{c,N} = \left[ \left( N+1 \right) !   {\epsilon_{T,N}  \over 2 } 
 \left( {4 \over \lambda} \right)^{(N+1)} \right]^{1/(4(N + 5/4))}.
\end{equation}
Note that the non-analytic dependence of $x_{c,N}$ on $\lambda$ arises
automatically from the imposition of  Eq.~(\ref{diff}) to satisfy the hypothesis
of the Lebesgue's theorem.

Remember that the only condition on $\epsilon_{T,N}$ to achieve convergence
of the sequence of integrals is to go to zero when $N \rightarrow \infty$
consistently with $x_{c,N} \rightarrow \infty$  in that limit. Choosing for 
example
\begin{equation}\label{prescrerr}
\epsilon_{T,N} = e^{- 4 N^{1/4}},
\end{equation}
we obtain asymptotically,
\begin{equation}\label{xcNassymp}
x_{c,N} \rightarrow  \left(  4 N / e \lambda \right)^{1/4}.
\end{equation}
This implies (through Eq.~(\ref{diff2})),
\begin{equation}\label{calcerr}
\epsilon_{c,N} \rightarrow e^{-  \left(  4 N / e \lambda \right)^{1/2} - N / e}.
\end{equation}
Equations~(\ref{prescrerr}) and~(\ref{calcerr}) show the exponential rate at
which the convergence of the sequence of integrals take place.

Clearly the form~(\ref{prescrerr}) for  $\epsilon_{T,N}$ is not unique, not even
the most efficient one, but enough to achieve convergence.

In the table~(\ref{table}) one can appreciate the numerical convergence
\begin{table}[t]\caption{ Integration over the small field configurations only 
produces a convergent series. In the last column  the improvement
over the perturbative values can be appreciated.\label{table}}
\begin{tabular}{|c|c|c|c|}
\hline
Order & Exact value ($\lambda=4/10$) & Conv. series & 
Pert. series  \\
\hline
 2 & 0.837043 & 0.803160 & 0.848839  \\
\hline
 4 & 0.837043 & 0.830264 & 0.854087  \\
\hline
 6 & 0.837043 & 0.835516 & 0.901897  \\
\hline
 8 & 0.837043 & 0.836667 & 1.316407  \\
\hline
 20 & 0.837043 & 0.837044 & 2.33755 $ 10^8$  \\
\hline
\end{tabular}
\end{table}
for  $\lambda = 4/10$.
\\

Up to now we have proved that the general strategy of section~\ref{sec:converg}
does, in fact, generate a convergent sequence towards $z (\lambda)$. 
However, the resulting integrals in~(\ref{exint3}) are not Gaussians, making
the applicability of the method  in Quantum Field Theory dubious, to say the
least. We will show now that the integrals of Eq~(\ref{exint3}) can be computed,
using Eq.~(\ref{win})  with  $u = (x/x_{c,N})^2$, calculating only Gaussian
integrals. The steps involved are
\begin{eqnarray}\label{calcnonGaussint1}
\int_{-x_{c,N}}^{x_{c,N}}  x^r e^{-x^2}  {\rm d}  x &=&
\int_{- \infty}^{\infty}  x^r e^{-x^2}  \lim_{M \rightarrow \infty} 
W(M,x,x_{c,N})  {\rm d}  x   \\   \label{calcnonGaussint2}
&=& \lim_{M \rightarrow \infty}  \int_{-\infty}^{\infty}  x^r e^{-x^2}  
W(M,x,x_{c,N})  {\rm d}  x    \\
&=& \lim_{M \rightarrow \infty} \sum_{n=0}^{M} {1 \over n!}  
\left( {M \over x_{c,N}^2} \right)^n  \int_{-\infty}^{\infty} 
e^{- \left( 1 + M / x_{c,N}^2 \right)  x^2}  x^{2n + r} {\rm d}  x  \nonumber \\
 \label{calcnonGaussint3}
\end{eqnarray}
The two properties of $W$ validate both equality~(\ref{calcnonGaussint1})
and (because of the uniformity of the convergence in 
$W$)~(\ref{calcnonGaussint2}).  In the last line~(\ref{calcnonGaussint3}) 
we just  make explicit the meaning of  (\ref{calcnonGaussint2}). So it is clear 
that these two properties are enough to prove the validity
of~(\ref{calcnonGaussint3}), where only Gaussian integrals are present. 
But it is a good exercise to find a {\it direct}  proof of it in the case at hand, 
where everything can be computed exactly. We do this next.

For $r$ odd the integrals vanish, so let's consider the case $r$ even, that is,
$r = 2 t$, for any integer $t$.

On the one hand we have
\begin{equation}\label{left}
\int_{-x_{c,N}}^{x_{c,N}}  x^{2t} e^{-x^2}  {\rm d}  x =
 \left( x_{c,N} \right)^{2t+1}  \sum_{k=0}^{\infty} {(-1)^k \over k!}  
 {\left( x_{c,N} \right)^{2k} \over \left( k+t+ 1/2 \right)} 
\end{equation}
where the necessary interchange between sum and integral to arrive to the
result is allowed due to the uniform convergence of the Taylor series of
$e^{- x^2}$ in the finite segment~$\left[ -x_{c,N}, x_{c,N} \right]$.

On the other hand,
\begin{eqnarray}\label{right1}
& & \lim_{M \rightarrow \infty} \sum_{n=0}^{M} {1 \over n!}  
\left( {M \over x_{c,N}^2} \right)^n  \int_{-\infty}^{\infty} 
e^{- \left( 1 + M / x_{c,N}^2 \right)  x^2}  x^{2 \left( n + t \right)} {\rm d}  x 
\qquad\qquad\qquad \\  \label{right2}
&=& \lim_{M \rightarrow \infty} \sum_{n=0}^{M} {1 \over n!}  \Gamma 
\left( n+t+ 1/2 \right)  \left( {x_{c,N}^2 \over  M} \right)^{t + 1/2}
\left(1 +  {x_{c,N}^2 \over M} \right)^{- \left( n +t + 1/2 \right)}   \\  \label{right3}
&=& \lim_{M \rightarrow \infty} \sum_{n=0}^{M} {1 \over n!}
\left( {x_{c,N}^2 \over  M} \right)^{t + 1/2}  \sum_{k=0}^{\infty}  {(-1)^k \over k!}
\Gamma \left( n+t+k+ 1/2 \right) \left( {x_{c,N}^2 \over  M} \right)^{k}  \\  
&=& \left( x_{c,N} \right)^{2t+1}  \sum_{k=0}^{\infty} {(-1)^k \over k!}  
 {\left( x_{c,N} \right)^{2k} \over \left( k+t+ 1/2 \right)} 
\left[ \lim_{M \rightarrow \infty} {(k+t+ 1/2) \over M^{(k+t+ 1/2)}}
\sum_{n=0}^{M}  { \Gamma \left( n+t+k+ 1/2 \right) \over n!}  \right] \nonumber \\
\label{right4}
\end{eqnarray}
In line~(\ref{right2}) we have used the equation
\begin{equation}\label{innt}
\int_{- \infty}^{\infty} x^{2n} e^{-p x^2} {\rm d} x  =  {\Gamma (n + 1/2)  \over 
p^{n+ 1/2}},
\end{equation}
in line (\ref{right3})  we have expanded the last term of  (\ref{right2}) in powers
of  $x_{c,N}^2 / M$ and carried out some cancellations, and finally  in 
(\ref{right4}) we have interchanged the $M \rightarrow \infty$ limit with
the infinite sum in $k$.

Comparing (\ref{left})  and (\ref{right4}), we see that the validity of Eq. (\ref{calcnonGaussint3})
depends on the validity of the equation
\begin{equation}\label{identity}
\lim_{M \rightarrow \infty} {(k+t+ 1/2) \over M^{(k+t+ 1/2)}}
\sum_{n=0}^{M}  { \Gamma \left( n+k+t+ 1/2 \right) \over n!}  = 1\quad
\forall \ {\rm integers} \ k, t > 0 
\end{equation}
That this identity holds for every integer $t$ and $k$ can be seen by
considering the following analytic function of the complex variable $z$:
\begin{equation}\label{analfunct}
O (z) \equiv \lim_{M \rightarrow \infty} {(1/z) \over M^{(1/z)}}
\sum_{n=0}^{M}  { \Gamma \left( n+1/z \right) \over \Gamma \left( n+1 \right)!}.
\end{equation}
If the identities (\ref{identity}) hold, this function must be identically 1, since for
$1/z_j = j + 1/2 $ with $j$ integer it reduces to them, and for ever increasing $j$,
we obtain a sequence accumulating at $z=0$ on which the function should be 1.

Conversely we will prove that $O(z)$ is indeed identically 1 as an analytic
function of $z$, proving in consequence the identities (\ref{identity}) for arbitrary
$t$ and $k$. Consider the sequence $1/z_j = j + 1$ for $j$ integer. This sequence
also accumulates at $z=0$, and for all its points we have
\begin{eqnarray}\label{proofident}
O \left( 1/ ( j + 1) \right) &=& \lim_{M \rightarrow \infty} {(j+1) \over M^{(j+1)}}
\sum_{n=0}^{M}  { \Gamma \left( n+ j+1 \right) \over \Gamma \left( n+1 \right)!}
\\
&=& \lim_{M \rightarrow \infty} {(j+1) \over M^{(j+1)}} \sum_{n=0}^{M}
\Pi_{i=1}^j (i+n)  \\
&=&  \lim_{M \rightarrow \infty} {(j+1) \over M^{(j+1)}} \left[ \sum_{n=0}^{M}
n^j + {\cal O} (n^{j-1}) \right] \\
&=&  \lim_{M \rightarrow \infty} {(j+1) \over M^{(j+1)}} 
\left[  { M^{(j+1)} \over (j+1)} +  {\cal O} (M^j) \right]  
\stackrel{M \rightarrow  \infty}{\longrightarrow} 1
\end{eqnarray}
Therefore $O(z) = 1$ for all $z$. This finishes the direct proof of
Eq.~(\ref{calcnonGaussint3}).

As was mentioned before, Eq.~(\ref{xcNasfunctN}), that was derived 
independently of any analyticity consideration, and only with the purpose
of  satisfying the hypothesis of Lebesgue's theorem, introduces a 
non-analyticity in the sequence of integrals of  $f_N'$ even for finite $N$. 
But even for the case where $x_{c,N}$ is fixed to a constant $a$, discussed
before, in which the limit $N \rightarrow \infty$ is taken first, and then $a$ is sent
to infinity, and therefore  the sequence is made out of truly analytic functions,
the convergence towards $z (\lambda)$ is perfectly compatible with analyticity
considerations.  The functions
$\pi^{-1/2} \int_{-a}^{a} {\rm d} x e^{ - \left(  x^2 +  {\lambda \over 4} x^4 \right)  }$
(the result of the $N \rightarrow \infty$ limit), are clearly analytic in $\lambda$.
But they converge to (in fact they define!) the nonanalytic function 
$z (\lambda)$ when $a \rightarrow \infty$. The limit of an infinite sequence of
analytic functions does not have to be analytic.

Another important issue is that the same method works also for the ``negative
mass case", where the Borel resummation method fails. Indeed, from the discussion
of this section it must be obvious that, with a proper scaling of  $x_{c,N}$, the
$f_N'$'s with negative quadratic part of the exponent also converges uniformly
towards $e^{ (x^2 -  {\lambda \over 4} x^4) }$ for $x$ in $\left[ - x_{c,N} , x_{c,N}
\right]$. Therefore, the sequence of integrals is also convergent.

\end{document}